\documentclass[twocolumn,prmaterials,superscriptaddress]{revtex4-1}
\usepackage{graphicx}
\usepackage{bbold}
\begin{document}

\title{Highly parallelizable electronic transport calculations \\ 
in periodic Rhodium and Copper nanostructures} 
\author{Shifeng Zhu}
\email{sfzhu@u.washington.edu}
\affiliation{Department of Physics, University of Washington, Seattle, Washington 98195, United States.}  
\author{Baruch Feldman}
\email{baruchf@u.washington.edu.  SZ and BF contributed equally to this work.} 
\affiliation{Department of Electrical and Computer Engineering, University of Washington, Seattle, Washington 98195, United States.} 
\author{Scott Dunham}
\affiliation{Department of Electrical and Computer Engineering, University of Washington, Seattle, Washington 98195, United States.} 
\affiliation{Department of Physics, University of Washington, Seattle, Washington 98195, United States.}  

\begin{abstract}
We extend the highly-parallelizable open-source electronic transport code TRANSEC \cite{transec,Transec-code} 
to perform real-space atomic-scale electronic transport calculations with periodic boundary conditions in the lateral dimensions.  We demonstrate the use of TRANSEC in periodic Cu and Rh bulk structures and in large periodic Rh point contacts, in preparation to perform calculations of reflection probability across Rh grain boundaries.  
\end{abstract} 

\maketitle

\section{Introduction}

  In order for semiconductor technology to continue scaling, the most local level of metal interconnects must scale with minimum feature size (5-7 nm in current technology nodes) to form logical circuits among the billions of individual transistor devices in an integrated circuit.  However, in metal wires with $\sim$10 nm line width, conductivity can degrade by as much as 90\% compared to the bulk metal \cite{Gall-Search}.  Thus, interconnect may in fact be the limiting factor in integrated circuit delay times and power consumption \cite{ITRS-2007}.  
  
	Key causes of this conductivity degradation, also known as the size effect, include scattering from grain boundaries, surface roughness, and wire-liner interfaces \cite{SurfScatter, Cu-GBs-Intel, Liner}.  
In this limit, resistivity $\rho$ is augmented over its bulk value $\rho_0$ by an amount scaling as the product of $\rho_0$ and bulk mean free path $\lambda$ divided by line width $L$: 
\[
\rho(L) - \rho_0 \; \sim \; \frac{\rho_0 \lambda}{L} \; .  
\]  
The product $\rho_0 \lambda$ has therefore been used to indicate which materials are promising candidates to replace copper in future nanoscale interconnects.  Previously, rhodium was identified as one of the most promising such materials \cite{Gall-MFP-promising-materials, Shifeng-mode-counting, Gall-Search}.  

In this article, we extend the highly-parallelizable open-source electronic transport code TRANSEC, based on the Density Functional Theory (DFT) code PARSEC \cite{transec, Transec-partitions, Transec-code, parsec-review, parsec-PBCs, parsec-code}, to perform real-space atomic-scale electronic transport calculations with periodic boundary conditions in the lateral dimensions (Sec.~\ref{sec:method-PBC}).  We then describe a mode counting method we use to validate our calculations (Sec.~\ref{sec:mode-counting}), and demonstrate the use of TRANSEC in periodic Rh and Cu structures (Sections \ref{sec:Cu-results} and \ref{sec:Rh-results}).  
In a subsequent work, we will present TRANSEC calculations of reflection probability across Rh grain boundaries, 
using the absorbing boundary condition parameters developed here to represent bulk Rh electrodes.

TRANSEC and PARSEC use a real-space grid to represent the Kohn-Sham (KS) DFT orbitals.  
The advantages of real-space calculations include an extremely sparse Hamiltonian, leading the computation to parallelize very efficiently.   
As described previously \cite{transec, Transec-partitions}, this allows real-space calculations to take better advantage of parallel computing resources in order to solve large, computationally intensive problems 
such as the ones studied here.  
In addition, real-space calculations allow straightforward convergence of the DFT ``basis'' set. 
Finally, real-space calculations naturally handle periodic, non-periodic, or mixed boundary conditions, as demonstrated in this work.  

TRANSEC uses absorbing boundary conditions (ABCs; also referred to as complex absorbing potentials, or CAPs) to represent semi-infinite electrodes at the two edges of the transport simulation cell.  The use of ABCs in the ``wide-band limit approximation'' avoids the computational burden of independent self-energy calculations for each energy $E$, making it possible to compute a dense transmission curve $T(E)$ in constant time \cite{transec}.   

\section{Method: real-space electronic transport calculations in periodic structures \label{sec:method-PBC}}

\subsection{Preliminaries and notation}

Let the Kohn-Sham effective potential, $V_{KS}(x,y,z)$, be periodic in the $xy$-plane with lattice vectors $\vec{a}$ and $\vec{b}$, and let $z$ be the transport direction.  
In this section, we describe the formalism for two-dimensional periodicity, or ``slab'' geometry, but the extension to the case of one-dimensional periodicity is straightforward.  
Note that since $\vec{a}$ and $\vec{b}$ span the $xy$-plane, 
$V_{KS}(x,y,z)$ is also at least approximately periodic in the $x$ and $y$ directions. 

Then Bloch's theorem tells us that the KS eigenstates $\phi$ can be chosen to have the Bloch form 
\begin{equation}
\phi(\vec{x}) = \phi(x,y,z) = u(\vec{x}) \: e^{i (\vec{k} \cdot \vec{x})} \; = \; u(x_a,x_b,z) \: e^{i (k_a x_a + k_b x_b)} \; ,
\label{eq:Bloch}
\end{equation}
where $\vec{k} \perp \hat{z}$, 
$x_{a,b}$ and $k_{a,b}$ are the components of $\vec{x}$ and $\vec{k}$, respectively, along the two lattice vectors, 
and 
$u$ is the periodic part of $\phi$, with periodicity $L_a = |\vec{a}|$ and $L_b = |\vec{b}|$ in $x_a$ and $x_b$,  
respectively.  

As always in TRANSEC \cite{transec}, we seek to evaluate the transmission function 
\[
T(E) = \mathrm{Tr}\{ G(E) \: \Gamma_R \: G^*(E) \: \Gamma_L \},
\] with $\Gamma_L$ and $\Gamma_R$ the ABCs in the left and right electrodes, respectively, which we take to have Gaussian form: 
\begin{equation} 
\Gamma_{L, R}(x,y,z) = \Gamma_0 \: e^{- {(z-z_{L, R})^2} / 2 \sigma^2 }  \; . 
\label{eq:CAP-Gaussian} 
\end{equation}
Here $\Gamma_0$ and $\sigma$ are the CAP strength and standard deviation, respectively, $\Gamma \equiv \Gamma_L + \Gamma_R$,   
\[
G(E) = \left[ E \mathbb{1} - H_{KS} + i \Gamma \right]^{-1}  
\]
is the retarded Green's function, $G^* = G^\dagger$ is the advanced Green's function, and $H_{KS}$ is the KS Hamiltonian.  

Following Ref.~\cite{transec}, we wish to carry out the trace in the diagonal basis of $G^{-1}$, i.e.~the basis of right eigenvectors $U$ of $(H_{KS} - i \Gamma)$.  
The eigenvectors $U$ can be chosen with the Bloch form (\ref{eq:Bloch}), and therefore each individual eigenvector $U_{n, \vec{k}}$ can be identified by a band index $n$ and Bloch wavevector $\vec{k}$. 
Accordingly, the formula for transmission becomes: 
\begin{equation}
T(E) = \mathrm{Tr}\{ \tilde{G}(E) \: \tilde{\Gamma}_R \: \tilde{G}^*(E) \: \tilde{\Gamma}_L \}, 
\label{eq:transmission-eigenbasis}
\end{equation} 
where 
\begin{eqnarray}
\nonumber \tilde{\Gamma}_{L} \equiv \: U^{\dagger} \: \Gamma_{L} \: U \: , \\
\tilde{\Gamma}_{R} \equiv \: U^{T} \: \Gamma_{R} \: U^* \; ,  
\label{eq:Gamma-tilde}
\end{eqnarray}  
\begin{equation}
\tilde{G}(E) \; \equiv \; U^{T} \: G(E) \: U = \: \mathrm{diag} \{ \: 1/\left( E - \epsilon_{n, \: \vec{k}} \right) \: \} \: , \;
\label{eq:G-diag}
\end{equation}  
and $\epsilon_{n, \: \vec{k}}$ is an eigenvalue of $(H_{KS} - i \Gamma)$.  
Note that $(H_{KS} - i \Gamma)$ is complex symmetric, so $U$ is complex orthogonal, $U^{-1} = U^T$, and the individual vectors $U_{n \vec{k}}$ are biorthogonal \cite{transec}.  

The CAPs (\ref{eq:CAP-Gaussian}) decay and are therefore clearly non-periodic in the $z$-direction, but they are in fact {\em constant} in the $xy$-plane.  Therefore, the transformed CAPs $\tilde{\Gamma}$ will be diagonal in the subspaces of $U$ with constant $\vec{k}$, while $\tilde{G}$ remains fully diagonal as in Eq.~(\ref{eq:G-diag}) even with PBCs.  So we need not consider the possibility that $\vec{k}$ complicates 
Eq.~(\ref{eq:transmission-eigenbasis}), 
and the 
formalism 
of Ref.~\cite{transec} 
can be adapted directly to the case with PBCs.  

Evidently, 
the main adaptation required to introduce PBCs is carrying out the basis change (\ref{eq:Gamma-tilde}) of the CAPs 
with 
an eigenbasis $U$ having Bloch form (\ref{eq:Bloch}).  
Formally, this transformation can be expressed as: 
\begin{eqnarray}
\tilde{\Gamma} = \int_\infty u'^{*}(\vec{x}) \: u(\vec{x}) \; e^{i ((\vec{k} - \vec{k'}) \cdot \vec{x})} \: \Gamma(\vec{x}) \: d^3 \vec{x} \nonumber  \\  
= \; \sum_{n,m} e^{i ((k_a - k'_a) n L_a + (k_b - k'_b) m L_b)} \frac{\vec{a} \times \vec{b}}{L_a L_b} \nonumber \\ 
\int\limits_{-\infty}^\infty \int\limits_0^{L_b} \int\limits_0^{L_a} \bar{u}'^{*}(\vec{x}) \: 
\bar{u}(\vec{x}) \; e^{i ((\vec{k} - \vec{k'}) \cdot \vec{x})} \: \Gamma(z) \: dx_a \: dx_b \;  dz \nonumber \\ 
 = \; \frac{\delta^2(\vec{k} - \vec{k'}) }{A} \int 
\bar{u}'^{*}(\vec{x}) \: 
\bar{u}(\vec{x}) \: \Gamma(\vec{x}) \: d^3 \vec{x} \; , \ 
\label{eq:transform-CAP}
\end{eqnarray}
where $A$ is the Born-von Karman cross-sectional area and $\bar{u}$ is the restriction of $u$ to the periodic cell, 
\[
0 \leq x_a < L_a \:, \; 0 \leq x_b < L_b \: .  
\]
The key observation 
is that we can transform $\Gamma_{L, R}$ using only the periodic portions $\bar{u}$ of the eigenbasis restricted to the periodic cell, and in fact do not need to keep track of the $\vec{k}$-dependent complex exponentials. 

As 
in Ref.~\cite{transec}, after basis change, the CAPs (\ref{eq:transform-CAP}) are no longer expected to be diagonal between different bands $n$ and $m$, so in Eq.~(\ref{eq:transmission-eigenbasis}) we must multiply the blocks of $\tilde{\Gamma}_{L, R}$ for each subspacesY of fixed $\vec{k}$.  But each matrix inside the trace in Eq.~(\ref{eq:transmission-eigenbasis}) is at least diagonal over $\vec{k}$, 
so the trace over $\vec{k}$ merely entails performing an independent multiplication and subspace trace for each sampled value of $\vec{k}$.  
As a result, $k$-space integration can be performed by computing the transmission $T_k(E)$ for each sampled $k$-point, and averaging $T_k(E)$ over the sampled $k$-points to estimate $T(E)$,
as is generally the case for $k$-averaged quantities 
\cite{Martin-e-structure}.  

We therefore express the Bloch state (\ref{eq:Bloch}) as a tensor product 
\begin{equation}
|U_{n,\vec{k}}\rangle = |u_{n,\vec{k}}\rangle \otimes |\vec{k} \rangle.  
\end{equation}
Here the tensor product is understood as connecting two disjoint parts of Hilbert space, in the sense that 
$|\vec{k} \rangle$ has $\vec{k}$ lying within the first Brillouin Zone (1BZ), whereas the periodic function $u_{n \vec{k}}$ can be Fourier expanded into reciprocal lattice vectors {\em outside} the 1BZ \cite{Ashcroft-Mermin}.  
We likewise decompose $H_{KS}$ into a sum of tensor products of parts $H_{\vec{k}}$ that operate only on $|u_{n,\vec{k}}\rangle$, and projection operators that select the $|\vec{k} \rangle$ component:  
\begin{equation}
\label{eq:H-projection-kpts}
H_{KS} = \frac{A}{(2 \pi)^2} \: \int_{\vec{k} \: \in \: 1BZ} H_{\vec{k}} \otimes |\vec{k} \rangle \langle \vec{k}| \; d^2{\vec{k}} \; , 
\end{equation}

where 
\begin{equation}
H_{\vec{k}} \equiv V_{KS}(\vec{x}) + \frac{\hbar^2}{2m}(\nabla^2 + k^2 - 2 i \vec{k} \cdot \vec{\nabla}) \: , 
\label{eq:H_k}
\end{equation} 
within the periodic cell \cite{Ashcroft-Mermin}.

\subsection{Transmission within a subspace of constant $\vec{k}$ \label{sec:transmission-k-point}}

We wish to substitute 
Eq.~(\ref{eq:transform-CAP}) into (\ref{eq:transmission-eigenbasis}), so as to express $T(E)$ in terms of $|u_{n \vec{k}}\rangle$.  
As in Ref.~\cite{transec}, our main computational problem is partial diagonalization 
of $G^{-1}$ in the neighborhood of the Fermi energy, $E_F$.  In the present work, this problem is modified by restricting $H_{KS}$ to each respective sampled 
$\vec{k}$-point, as in 
(\ref{eq:H_k}) and the discussion above.    

However, note that the transmission formulas of Ref.~\cite{transec}, and its orthonormalization convention for the eigenvectors $U$, assume $G^{-1}$ to be complex symmetric. 
By contrast, 
Eq.~(\ref{eq:H_k}) is Hermitian but not real (not time-reversible), so $H_{\vec{k}}$ is not symmetric, $H_{\vec{k}} - i \Gamma$ is \emph{not} complex symmetric, and the $|u_{n \vec{k}}\rangle$ are {\em not} biorthogonal.  
We seek to restore complex symmetry  
by rearranging Eq.~(\ref{eq:H-projection-kpts}): 
\begin{eqnarray}
\nonumber
H_{KS} = \\
\nonumber \frac{A}{(2 \pi)^2} \: \int\limits_{1BZ_+ } 
H_{\vec{k}} \otimes |\vec{k} \rangle \langle \vec{k}| +  H_{-\vec{k}} \otimes |-\vec{k} \rangle \langle -\vec{k}| \; d^2{\vec{k}} \\
= \frac{A}{(2 \pi)^2} \: \int\limits_{1BZ_+} 
2 \: \mathrm{Re}\left\{ H_{\vec{k}} \otimes |\vec{k} \rangle \langle \vec{k}| \right\} \; d^2{\vec{k}} .
\label{eq:H-restore-reality}
\end{eqnarray}
Here the integration domain is 
half the 1BZ: 
\[
1BZ_+ \; \equiv \; 
\vec{k} \; \in \: \left\{ 1BZ \; | \; k_a \geq 0 \right\} \; , 
\]
but we have suppressed special treatment of points like $\vec{k}=0$ for simplicity of presentation.  
Therefore, 
we consider 
\begin{equation}
G_{\vec{k}}^{-1}(E) \equiv E \mathbb{1} - 2 \: \mathrm{Re}\left\{ H_{\vec{k}} \otimes |\vec{k} \rangle \langle \vec{k}| \right\} + i \Gamma \; ,
\label{eq:def-Gk}
\end{equation}
which acts on both $|u_{\pm \vec{k}}\rangle$ for each sampled value of $\vec{k}$ in the integration domain of (\ref{eq:H-restore-reality}), and by construction \emph{is} 
complex symmetric.  

With the rearrangement (\ref{eq:H-restore-reality}) and the definition (\ref{eq:def-Gk}), we can obtain a biorthogonal basis of eigenvectors for $G_{\vec{k}}^{-1}$, as in Ref.~\cite{transec}.  By inspection, we propose 
these 
have the form 
\begin{eqnarray}
|v_{\vec{k}}\rangle_+  = \frac{1}{\sqrt{2}} \left( |u_{\vec{k}}\rangle \otimes |\vec{k} \rangle  \; + \; |u_{-\vec{k}}\rangle \otimes |-\vec{k} \rangle \right), \nonumber \\  
i |v_{\vec{k}}\rangle_-  = \frac{1}{\sqrt{2}} \left( |u_{\vec{k}}\rangle \otimes |\vec{k} \rangle  \; - \; |u_{-{\vec{k}}}\rangle \otimes |-\vec{k} \rangle \right).  
\label{eq:form-real-evectors}
\end{eqnarray}
Note that $| v_{n k} \rangle_-$ 
is undefined for the $\Gamma$ point, $\vec{k} = 0$, since here $\vec{k} = -\vec{k}$.  

To validate the form of (\ref{eq:form-real-evectors}) as eigenvectors, we must confirm that $|u_{\pm \vec{k}} \rangle$ lie in the same eigenspace, i.e.~share the same eigenvalue.  
We start by considering the eigenvectors $|u_{\pm \vec{k}}^0 \rangle$ and eigenvalues $\epsilon_{\vec{k}}^0$ of $H_{\pm \vec{k}}$ without $i\Gamma$.  
These are, of course, the KS-DFT eigenpairs of the one- or two-dimensional periodic DFT calculation \emph{without} CAPs.  
Because $H_{\vec{k}}$ is Hermitian and $H_{-\vec{k}} = H_{\vec{k}}^{\; *}$ in the real-space basis employed by TRANSEC, time-reversal symmetry or complex conjugation of the eigenvalue equation show that $|u_{-\vec{k}}^0\rangle \propto |u_{\vec{k}}^{0 *}\rangle$ and $\epsilon_{-\vec{k}}^0 = \epsilon_{\vec{k}}^{0\; *} = \epsilon_{\vec{k}}^0$ .  
Thus, 
both $H_{\pm \vec{k}}$ have the same eigenvalue $\epsilon_{\vec{k}}^0$ , albeit different eigenvectors, 
and the linear combinations $|v_{\vec{k}}^{\; 0}\rangle_\pm$ 
remain eigenvectors of  $\mathrm{Re}\left\{ H_{\vec{k}} \otimes |\vec{k} \rangle \langle \vec{k}| \right\}$ .

\subsection{With $R_{z,\pi}$ symmetry \label{sec:with-rzpi-symm}}

The time-reversal symmetry we have used to relate $|u_{-\vec{k}}^0\rangle$ to $|u_{\vec{k}}^{0 *}\rangle$ is present in equilibrium KS-DFT, but broken in the electronic transport problem.  However, although the addition of the CAPs $i \Gamma$ breaks time-reversal symmetry of the Hamiltonian, $\epsilon_{\pm \vec{k}}$ remain equal.  
This is manifestly true 
when $V_{KS}$ respects 
$C_2$ symmetry about the transport axis, i.e.~when $V_{KS}$ is symmetric under rotation by $\pi$ about $z$, or equivalently 2-D inversion in the $xy$-plane, which maps between ${\pm \vec{k}}$.  For clarity, we will refer to this as the $R_{z,\pi}$ symmetry operation, rather than by the usual 
point-group symbol $C_2$.  
With $R_{z,\pi}$ symmetry, $|u_{+\vec{k}}\rangle$ and $|u_{-\vec{k}}\rangle \propto R_{z,\pi} |u_{+\vec{k}}\rangle$ 
are both eigenvectors of $H_{\pm\vec{k}} \; + \: i \Gamma$ with the same eigenvalue $\epsilon_{\vec{k}}$, so $|v_{\vec{k}}\rangle_\pm$ continue to be eigenvectors of Eq.~(\ref{eq:def-Gk}). 
 
Now that we have validated the form of the vectors $|v_{\vec{k}}\rangle_\pm$ for the case with $R_{z,\pi}$ symmetry, we wish to check their normalization in preparation to apply the transformations (\ref{eq:Gamma-tilde}).  
Different $|k\rangle$'s remain orthogonal to each other, as in Eq.~(\ref{eq:transform-CAP}) above.  So normalizing (\ref{eq:form-real-evectors}): 
\begin{eqnarray}
\label{eq:normalize-real-evectors}
|v_k\rangle_\pm \cdot |v_k\rangle_\pm  = + \frac{ |u_k\rangle \cdot |u_{-k}\rangle + |u_{-k}\rangle \cdot |u_{k}\rangle}{2} \nonumber \\
= |u_{-k}\rangle \cdot |u_k\rangle  
= \langle u_{-k}^* | u_k\rangle \; , 
\end{eqnarray}
while $|v_k\rangle_\pm \cdot |v_k\rangle_\mp  = 0$.  Here we use notation where a dot product represents the biorthogonal scalar product of two KS states: $|a\rangle \cdot |b\rangle \equiv \int a(\vec{x}) \: b(\vec{x}) \; d^3 \vec{x}$, while the bracket represents the standard inner product:  $\langle a | b \rangle = |a\rangle^* \cdot |b\rangle$. 
Thus, if $|u_{+\vec{k}}\rangle$ is normalized against $|u_{-\vec{k}}\rangle$ via the biorthogonal scalar product, the $|v_{\vec{k}}\rangle_\pm$ are properly normalized also.  

By contrast, 
transforming $\Gamma$ to $\tilde{\Gamma}$ proceeds according to Eqs.~(\ref{eq:Gamma-tilde}), and can mix different bands $n$ and $m$: 
\begin{eqnarray}
\label{eq:transform-CAP-2}
\tilde{\Gamma}_{L,n k \pm, \: m k' \pm} = \: _{\pm}\langle v_{n k}| \Gamma_L | v_{m k'} \rangle_{\pm} \nonumber \\
= \frac{\delta^2(k - k')}{2 A} \left\{ \langle u_{n k}| \Gamma_L | u_{m k} \rangle +  \langle u_{n -k}| \Gamma_L | u_{m -k} \rangle \right\} \nonumber 
\\
= \frac{\delta^2(k - k')}{A} \langle u_{n k}| \Gamma_L | u_{m k} \rangle , \\ 
\nonumber \tilde{\Gamma}_{R,n k \pm, \: m k' \pm} = |v_{n k}\rangle \cdot \Gamma_R | v_{m k'}^* \rangle \\
= \nonumber \: _{\pm}\langle v_{n k}^*| \Gamma_R | v_{m k'}^* \rangle_{\pm} = \frac{\delta^2(k - k')}{A} \langle u_{n k}^*| \Gamma_R | u_{m k}^* \rangle \\  
= \frac{\delta^2(k - k')}{A} \langle u_{m k} | \Gamma_R | u_{n k} \rangle .  
\label{eq:transform-CAP-3}
\end{eqnarray}
Here we have reduced each $\Gamma_{L,R}$ expression to a single term by making use of 
$|u_{-\vec{k}}\rangle \propto R_{z,\pi} |u_{+\vec{k}}\rangle$ , which is valid only when 
$R_{z,\pi}$ symmetry is present, and of $\left[ R_{z,\pi}, \; \Gamma_{L,R} \right] \: = \: 0$, which is always true.  
Substituting these expressions into (\ref{eq:transmission-eigenbasis}), and following Ref.~\cite{transec}, we see that 
\begin{equation}
T_k(E) = \sum_{m,n}^N \frac{\tilde{\Gamma}_{L, m k, n k} \: \tilde{\Gamma}_{R, n k, m k}}{(E-\epsilon_{n k})(E-\epsilon^*_{m k})} \; .  
\label{eq:T-tilde-realsp}
\end{equation}
In practice, the summation can be restricted to a small fraction $p$ of the total of $N$ bands, having $\epsilon_{m k}$ and $\epsilon_{n k}$ in the neighborhood of $E_F$ \cite{transec}.  

\subsection{Lacking $R_{z,\pi}$ symmetry \label{sec:no-rzpi-symm}}

$R_{z,\pi}$ symmetry is absent from many important atomic systems, for example most non-twin grain boundaries we wish to study, and even some bulk orientations. 
Without $R_{z,\pi}$ symmetry, we have no definite relationship between $|u_{\pm\vec{k}}\rangle$, 
and consequently, our method must expend additional computing time to solve for both $|u_{\pm\vec{k}}\rangle$, 
as described in the Appendix.  
Still, the form (\ref{eq:form-real-evectors}) of the biorthogonal eigenvectors of $G_{\vec{k}}^{-1}$ remains valid, as we now show by treating the decaying CAPs $i \Gamma$ as a small perturbation \cite{transec} in Eq.~(\ref{eq:def-Gk}).  
Moreover, the explicit computations presented in Sections \ref{sec:Cu-results} and \ref{sec:Rh-results} below 
serve to 
demonstrate that $\epsilon_{\pm \vec{k}}$ are equal in practice. 

We argue schematically that the eigenvalue corrections $\epsilon_{\pm \vec{k}}^M$ are equal to all orders $M$ in perturbation theory (PT) in $i \Gamma$, and explicitly confirm this for the first- and second-order corrections.  
All perturbative corrections to both the eigenvalues and vectors are 
proportional to products and powers of 
\[ 
\langle U_{n \vec{k}}^0 \: | \: i \Gamma \: | \: U_{m \vec{k}'}^0\rangle 
= \frac{\delta^2(\vec{k} - \vec{k}')}{A} \; \langle u_{n \vec{k}}^0 \: | \: i \Gamma \: | \: u_{m \vec{k}}^0\rangle \;, 
\]
since $\Gamma$ is uniform in the $xy$-plane, as shown in Eq.~(\ref{eq:transform-CAP}).  
Thus the perturbation $i \Gamma$ could cause the un-perturbed bands to mix, but \emph{not} different k-points, and so should not affect the form of (\ref{eq:form-real-evectors}). 
Note these brackets, and the 
PT expressions they represent, should be well-defined since the states inside the brackets are un-perturbed, i.e.~eigenvectors of the Hermitian operators $H_{\pm k}$.  

Moreover, the perturbation $i \Gamma$ is anti-Hermitian (since it is purely imaginary and diagonal in the real-space basis): 
\begin{equation} 
\langle u_{n \vec{k}}^0 \: | \: i \Gamma \: | \: u_{m \vec{k}}^0\rangle = -\langle u_{m \vec{k}}^0 \: | \: i \Gamma \: | \: u_{n \vec{k}}^0\rangle^*  \; = \langle u_{m -\vec{k}}^0 \: | \: i \Gamma \: | \: u_{n -\vec{k}}^0\rangle   \; ,  
\label{eq:PT-correction-time-reversal}
\end{equation} 
as can also be seen explicitly from complex conjugation of Eq.~(\ref{eq:transform-CAP}).  Here the last equality follows from  $|u_{-\vec{k}}^0\rangle \propto |u_{\vec{k}}^{0 *}\rangle$, which is true for the {\em un-perturbed} states, and from $-(i \Gamma)^* = +i \Gamma$.   

In particular, the $M$th-order PT eigenvalue correction $\epsilon_{n, \vec{k}}^M$ depends on a sum of fractions like 
\[
\sum\limits_{m_1, \: \ldots, \: m_{M-1} \: \neq \: n} 
\frac{\langle u_{n \vec{k}}^0 \: | \: i \Gamma \: | \: u_{m_1 , \: \vec{k}}^0\rangle 
\; \ldots \; 
\langle u_{m_{M-1} , \: \vec{k}}^0 \: | \: i \Gamma \: | \: u_{n \vec{k}}^0\rangle}
{\left(\epsilon_{n, \vec{k}}^0 \: - \: \epsilon_{m_1, \: \vec{k}}^0 \right)
\; \ldots \; 
\left(\epsilon_{n, \vec{k}}^0 \: - \: \epsilon_{m_{M-1}, \: \vec{k}}^0 \right) } \; , 
\]
where there are $M$ brackets 
in the numerator and $(M-1)$ factors in the denominator.  
Eq.~(\ref{eq:PT-correction-time-reversal}) implies that motion reversal $\vec{k} \to -\vec{k}$ transposes each bracket in the numerator, which rearranges the summation indices $m_1, \: \ldots, \: m_{M-1}$, but leaves the sum overall unchanged.  
Hence, we argue that to each order $M$ in PT, $\epsilon_{n, -\vec{k}}^M \; = \; \epsilon_{n \vec{k}}^M$ .  
For example, the first-order eigenvalue correction is equal for $\pm \vec{k}$: 
\[
\nonumber \epsilon_{n, -\vec{k}}^1 = \langle u_{n -\vec{k}}^0 \: | \: i \Gamma|u_{n -\vec{k}}^0\rangle = \langle u_{n \vec{k}}^0 \: | \: i \Gamma \: | \: u_{n \vec{k}}^0\rangle = \epsilon_{n \vec{k}}^1 \; . 
\]
Likewise, the second-order eigenvalue correction $\epsilon_{n \vec{k}}^2 $ is a sum over $m \neq n$ of terms proportional to $|\langle u_{m \vec{k}}^0 | i \Gamma | u_{n \vec{k}}^0 \rangle|^2$, so Eq.~(\ref{eq:PT-correction-time-reversal}) shows that $\epsilon_{n, -\vec{k}}^2 = \epsilon_{n \vec{k}}^2$.  
As a result, $| u_{\pm \vec{k}} \rangle$ continue to be degenerate, and the linear combinations (\ref{eq:form-real-evectors}) remain eigenvectors of (\ref{eq:def-Gk}). 

We also note 
that the un-perturbed states $| u_{n, \pm k}^0 \rangle$ are complex conjugates of each other, but have the same real eigenvalue $\epsilon_{n \vec{k}}^0$ due to time-reversal symmetry.  However, the {\em perturbed} states 
are not complex conjugates since the $i \Gamma$ term in (\ref{eq:def-Gk}) retains its sign when $\vec{k}$ is reversed.  
Evidently the perturbed $|v_{n k} \rangle$ and 
$|v_{n k} \rangle^*$ would instead be eigenvectors of $G_k^{-1}$ and $G_k^{-1 \: *}$, respectively (corresponding to the retarded and advanced Green's functions), with conjugate eigenvalues.

When $R_{z,\pi}$ symmetry is absent, 
the trace formula (\ref{eq:T-tilde-realsp}) 
must be modified to include {\em cross terms} between $| v_{n k} \rangle_+$ and $| v_{m k'} \rangle_-$  
having schematic form 
$_{\pm}\langle v| \Gamma | v \rangle_{\mp}$ , which are antisymmetric under interchange of $+$ and $-$.  
To this end, we consider a formula for such cross contributions in the CAP basis change: 
\begin{eqnarray}
\label{eq:transform-CAP-4}
&\tilde{\Gamma}_{L,n k \pm, \: m k' \mp} = \: _{\pm}\langle v_{n k} | \Gamma_L | v_{m k'} \rangle_{\mp} \\
&= \pm \frac{\delta^2(k - k')}{2 A i} \left\{ \langle u_{n k} | \Gamma_L | u_{m k} \rangle - \langle u_{n, -k} | \Gamma_L | u_{m, -k} \rangle \right\} \; , \nonumber \\
\label{eq:transform-CAP-5}
&\tilde{\Gamma}_{R,n k \pm, \: m k' \mp} = \: _{\pm}\langle v_{n k} | \Gamma_R | v_{m k'} \rangle_{\mp} \\
&= \pm \frac{\delta^2(k - k')}{2 A i} \left\{ \langle u_{n k}^* | \Gamma_R | u_{m k}^* \rangle - \langle u_{n, -k}^* | \Gamma_R | u_{m, -k}^* \rangle \right\} \nonumber \\ 
&= \pm \frac{\delta^2(k - k')}{2 A i} \left\{ \langle u_{m k} | \Gamma_R | u_{n k} \rangle - \langle u_{m, -k} | \Gamma_R | u_{n, -k} \rangle \right\} \: \nonumber . \; 
\end{eqnarray} 
With $R_{z,\pi}$ symmetry, $|u_{-\vec{k}}\rangle \propto R_{z,\pi} |u_{+\vec{k}}\rangle$, so Eqs.~(\ref{eq:transform-CAP-4}) and (\ref{eq:transform-CAP-5}) manifestly vanish, but without $R_{z,\pi}$ symmetry, we must generalize Eq.~(\ref{eq:T-tilde-realsp}) to include cross terms:  
\begin{equation}
T_k(E) = \sum_{m,n}^N \: \sum_{s, t = \pm 1} \: \frac{\tilde{\Gamma}_{L, m k s, n k t} \: \tilde{\Gamma}_{R, n k t, m k s}}{(E-\epsilon_{n k})(E-\epsilon^*_{m k})} \; .  
\label{eq:transform-no-symm}
\end{equation}

In TRANSEC, we   
explicitly compute $\tilde{\Gamma}_{m n}$ for $m \leq n$, and then 
require $\tilde{\Gamma}_{n m} \equiv \tilde{\Gamma}_{m n}^*$.  
We remark that the summand of (\ref{eq:transform-no-symm}) 
is symmetric under transposition of $s$ and $t$.  
In particular, 
the signs and factors of $i$ in $\tilde{\Gamma}_{L}$ cancel those in $\tilde{\Gamma}_{R}$, thus it is unnecessary to keep detailed track of whether these correspond to $\tilde{\Gamma}_{L}$ or $\tilde{\Gamma}_{R}$.  
Some other details of the implementation in TRANSEC of the methods described in this section are provided in the Appendix.

\section{Method: mode counting \label{sec:mode-counting}}

To validate the method of Sec.~\ref{sec:method-PBC}, 
we also make use of a mode-counting method to compute the ballistic conductance of bulk materials.  
The ballistic conductance $G_B$ of a bulk sample of cross-section $A$ is proportional to the number of current-carrying modes in the sample \cite{Datta}:  
 \begin{equation}
 \frac{G_B h}{2 e^2 A} = \frac{T_B\left(E_F\right)}{A} = \frac{M\left(E_F\right)}{A} = \frac{1}{(2 \pi)^2} \int{ \hat{n}_\perp \cdot \hat{z} \; \mathbf{d}^2 k_{||} } \: , 
\label{eq:cond-ballist}
 \end{equation}
where $G_B$ is ballistic conductance, $T_B(E)$ is ballistic transmission at energy $E$, $\hat{n_\perp} \propto 
\vec{\nabla}_k \: E$ is a unit vector normal to the Fermi surface, $\hat{z}$ is a unit vector conjugate to the transmission direction, $M(E)$ is the number of forward-moving modes with energy $E$, and the integration domain is the set of points on the Fermi surface with $\hat{n_\perp} \cdot \hat{z} > 0$. 
Note our convention that $T_B(E)$ and $M(E)$ are per-spin quantities, hence the factor of 2 in the denominator of the LHS.  

Our mode-counting method computes a discrete approximation to Eq.~(\ref{eq:cond-ballist}) 
based on a Fermi surface constructed from DFT.  
Our approach is to discretize the Brillouin Zone into tetrahedra, and find the differential element $\Delta k_{||} \approx d^2 k_{||}$ of the Fermi surface in each tetrahedron \cite{Gall-MFP-promising-materials, Shifeng-mode-counting}.  
The integration in Eq.~(\ref{eq:cond-ballist}) is thus transformed into a sum of $(\hat{n}_\perp \cdot \hat{z}) \; \Delta k_{||}$ inside the tetrahedra.  

Our calculation of electronic structure is performed with the Vienna Ab-initio Simulation Package (VASP) \cite{VASP-PRB, VASP-CMS} DFT code using the local density approximation (LDA).  For calculations of the bulk Cu and Rh conductances, we considered the primitive cell of the face-centered cubic (FCC) lattice, 
and employed a 48 $\times$ 48 $\times$ 48 Monkhorst-Pack $k$-point grid.  
When testing for convergence of the $k$-point sampling, we also confirmed that the surface integral Eq.~(\ref{eq:cond-ballist}) was converged.  

\begin{figure}
(a) \includegraphics[scale=0.45]{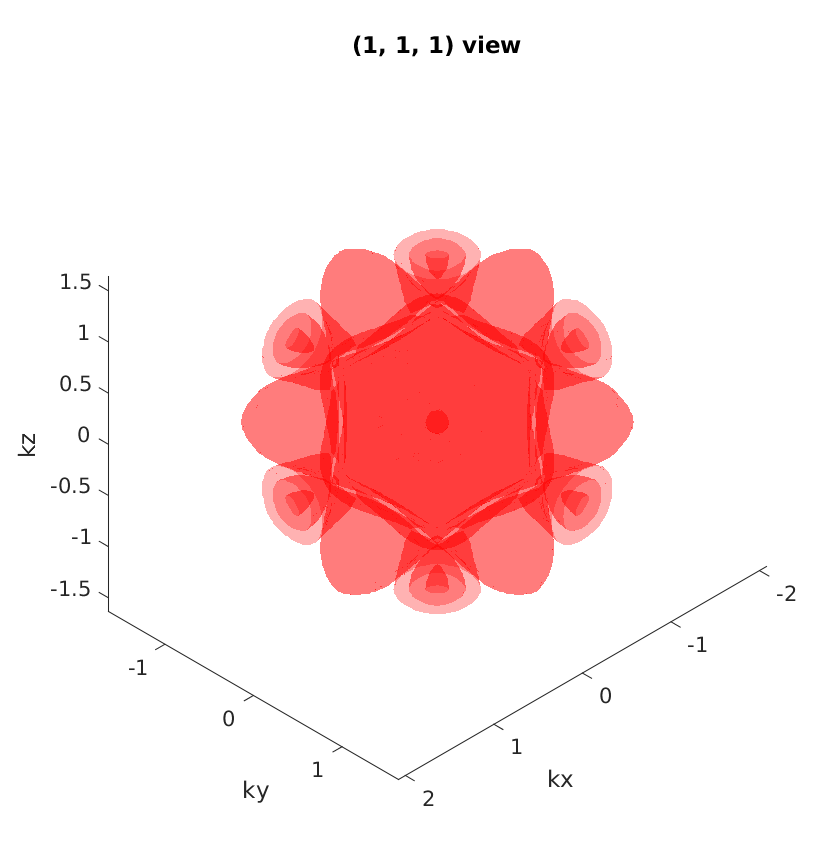} \\
(b) \includegraphics[scale=0.45]{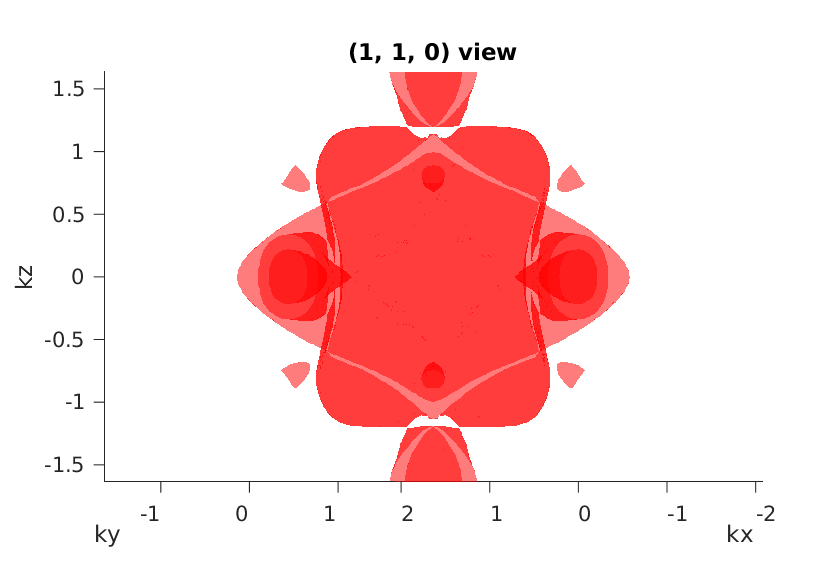} \\
\caption{Fermi surface projections for Rh in the (a) (111) and (b) (110) planes generated with our mode-counting method \cite{Shifeng-mode-counting}.  \label{fig:Fermi-surface-MC}}
\end{figure}

We used 
VASP to compute the KS energy levels on the Monkhorst-Pack grid 
in the Irreducible Brillouin Zone (BZ).  We considered tetrahedra formed from sets of four neighboring $k$-points covering the Irreducible BZ, and built the full 1BZ by applying symmetry operations on the Irreducible BZ. 
To find the Fermi surface within each tetrahedron, 
we interpolated the energy $E_n$ of each band $n$ linearly between the four vertices.  
The Fermi surface is then the surface of constant energy $\left\{ E_n(k) \equiv E_F, \; \forall n \right\}$ within the tetrahedron,  
iterating over bands 
$n$.  
The resulting Fermi surface projections for Rh in the 
(110) and (111) planes are shown in Figure \ref{fig:Fermi-surface-MC}. 

The Fermi velocity for band $n$ is given by 
\[
\vec{v}_{F,n} = \frac{1}{\hbar} \vec{\nabla}_k \: E_n \; . 
\]
We set $\hat{n}_\perp = \vec{v}_{F,n} / |\vec{v}_{F,n}|$ for the normal direction of the Fermi surface.   
This method is described in greater detail in Ref.~\cite{Shifeng-mode-counting}.  

\section{Results for periodic Cu structures with bulk electrodes \label{sec:Cu-results}} 

We now turn to demonstrating the methods of Sec.~\ref{sec:method-PBC} with bulk transmission calculations in transition metals, along the (100) and (111) transmission directions of the face-centered cubic (FCC) lattice.  In this Section we address Cu, and in Sec.~\ref{sec:Rh-results}, Rh.  
Note that the PARSEC and TRANSEC source codes implementing our methods, and example input and output from the simulations we performed, are available online \cite{Transec-code, parsec-code}.  

The (100) bulk calculations presented in Sections \ref{sec:bulk-Cu}, and \ref{sec:bulk-Rh} possess $R_{z,\pi}$ symmetry, whereas the (111) bulk, disorder, and point contact calculations presented in Sections \ref{sec:disorder-vacancy}, \ref{sec:bulk-Rh}, and \ref{sec:point-contacts} do not.  We also performed other validation tests with $R_{z,\pi}$ symmetry intentionally broken, and make extensive use of Sec.~\ref{sec:no-rzpi-symm} in our forthcoming work on Rh grain boundaries.  

\subsection{Bulk FCC Cu \label{sec:bulk-Cu}}

To validate the PBC method described in Sec.~\ref{sec:method-PBC}, we first computed ballistic conductance in bulk FCC Cu in the (100) orientation.  
For the (100) calculations here and in Sec.~\ref{sec:bulk-Rh} below, we used a tetragonal periodic simulation cell.  Specifically, the lattice vectors for the lateral PBCs are the primitive cell for the {\em two}-dimensional lattice of a (100) monolayer, pointing from the vertices of the cubic lattice to the face-centers of two adjacent cubic cells.  Thus, the two lateral cell axes are rotated so the basic repeating unit contains only two atoms, compared to four in the cubic unit cell.  Meanwhile, the transmission dimension aligns with the (001) direction of the usual FCC unit cell.  We chose this geometry because it is analytically simpler 
than the 3D primitive cell, having orthogonal lattice vectors, but has a smaller cross-section than the cubic unit cell so as to reduce the computational burden.   Of course, 
this choice of simulation cell orientation is not expected to affect any physical results of the calculation.  

The Cu lattice constant was chosen as $a_{Cu} =$ 6.77 $a_0$, the result of a lattice constant optimization of the bulk primitive cell.  Thus, the tetragonal simulation cell used 2-D PBCs with orthogonal lattice vectors towards two adjacent face centers, of length $a_{Cu} / \sqrt{2} =$ 4.79 $a_0$.  The calculation used 18 monolayers along the transport dimension, for a total length of 17 $a_{Cu} / 2 =$ 57.5 $a_0$ between the first and last layer.  In the transport dimension, we padded the cell length with an additional 8.8 $a_0$ of vacuum on each end to allow the outermost orbitals to decay, resulting in a total cell length of 75 $a_0$.

Our DFT calculation in PARSEC used LDA and 11 $\times$ 11 $k$-point sampling.  
We used a norm-conserving Troullier-Martins pseudopotential for Cu with electronic configuration of 3$d^{10}$4$s^1$4$p^0$ and $s$/$p$/$d$ cutoff radii of 2.05 / 2.30 / 2.05 $a_0$.  We used a grid spacing of $h =$ 0.37 $a_0$, after checking for convergence in the bulk primitive cell, so the simulation cell contained a total of $N =$ 29,100 grid points.  

For the transmission calculation, the results presented here 
used Gaussian CAPs, as in Eq.~(\ref{eq:CAP-Gaussian}), centered on the outermost Cu monolayers, with strength $\Gamma_0 =$ 100 mRy and standard deviation $\sigma =$ 8.5 $a_0$, providing contact regions of approximately 57.5 $a_0 \; /  (2 \cdot 8.5 \; a_0) = $ 3.4 standard deviations for each CAP to decay before the central region.  
We also tested for convergence of the contact length.  
In addition, we performed calculations using $\Gamma_0 =$ 77 mRy, $\sigma =$ 7.6 $a_0$ and $\Gamma_0 =$ 120 mRy, $\sigma =$ 9.2 $a_0$, and found the $T(E)$ results were very insensitive to these different choices, an indication the CAP parameters are valid \cite{transec}.  
We computed a fraction $p =$ 2\% of total complex eigenpairs \cite{transec}, and tested this fraction for convergence.  
Both 
DFT and transmission used an 11 $\times$ 11 Monkhorst-Pack grid, for a total of 61 $k$-points after application of $R_{z,\pi}$ symmetry.  

\begin{figure}
\includegraphics[scale=0.53]{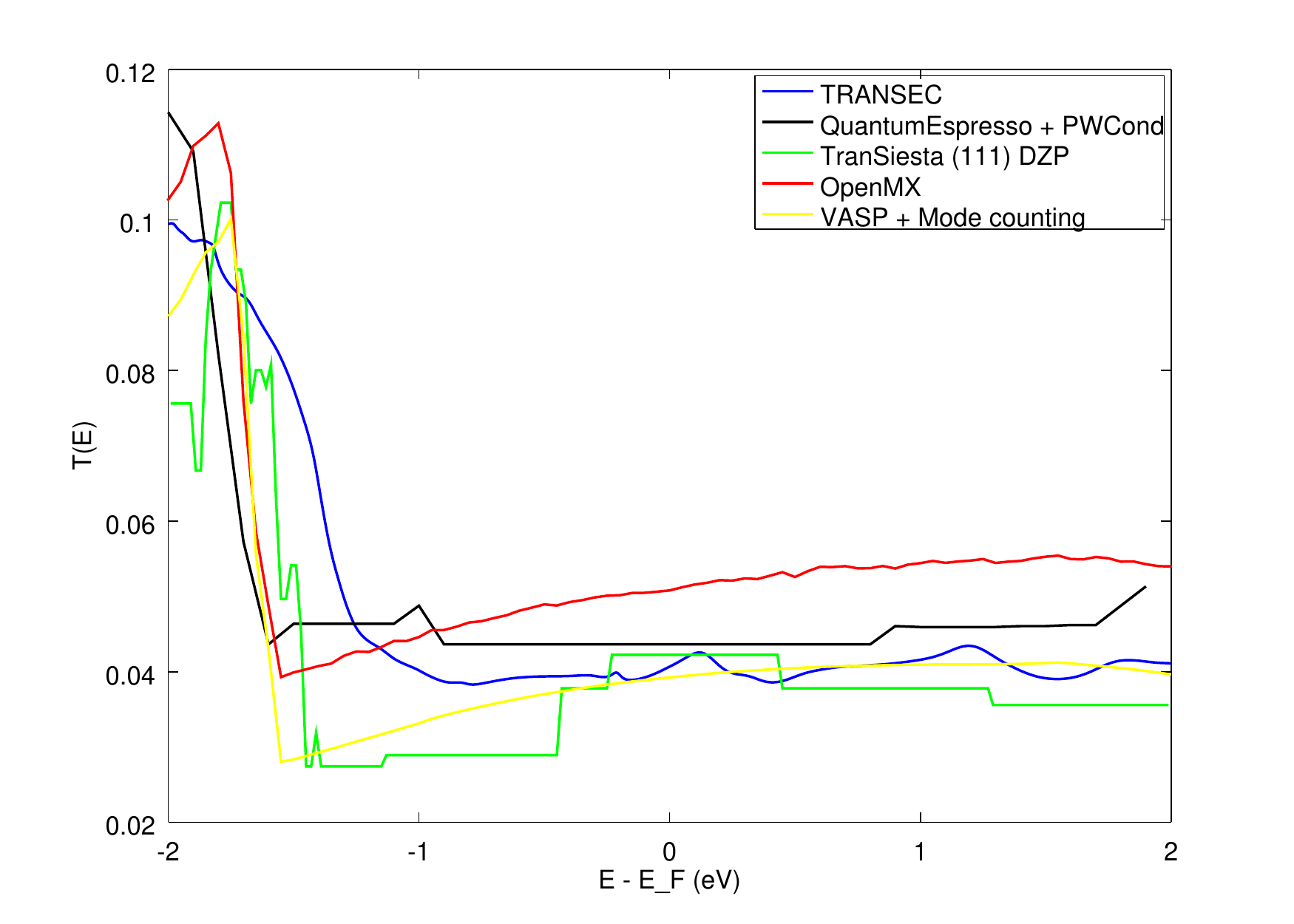} \\
\caption{Cu (100) bulk transmission per unit area, $T(E) / A$, as a function of energy $E$.  Also shown for comparison are results from OpenMX, PWCOND, Atomistix \cite{Cu-GBs-Intel}, TranSiesta, and our mode-counting method \cite{Shifeng-mode-counting}.  \label{fig:Cu-bulk-T}}
\end{figure}

Fig.~\ref{fig:Cu-bulk-T} shows the computed transmission $T(E)$ for bulk FCC Cu in the (100) orientation.  Also shown for comparison are results from OpenMX \cite{OMX-1, OMX-2}; PWCOND \cite{PWCond} using electronic structure from Quantum Espresso \cite{QE}; 
TranSiesta \cite{Brandbyge} using double-zeta polarized (DZP) orbitals; and our mode-counting method (see Sec.~\ref{sec:mode-counting}) 
using electronic structure from VASP.  
Note we performed the TranSiesta calculation for the (111) orientation, but $T(E)$ for this orientation is similar to $T(E)$ for (100).  
The other (100) calculations shown used the same tetragonal lattice vectors as the TRANSEC calculation.  Also note that PWCOND uses boundary value matching, TranSiesta uses self-energies to represent the electrodes, and OpenMX implements a mode-counting method similar to the one described in Sec.~\ref{sec:mode-counting} for ballistic calculations.  Therefore, none of these calculations used CAPs, so all used shorter simulation cells in the transmission dimension than our TRANSEC calculations.  As shown, these calculations agree 
with the TRANSEC calculation.  

We find a ballistic conductance per unit cross-sectional area of $T_B/A \; = \; T(E_F)/A \;=$ 0.93 $/ \; (4.79 \; a_0)^2 = $ 0.041 $a_0^{-2}$.  This finding is also in agreement with a simple Sommerfeld model \cite{Ashcroft-Mermin} assuming the Cu Fermi surface is modeled by a spherical shell of radius the Fermi wavevector $k_F$, and that Cu has one conductance electron per atom. 
Thus, in this model, the conduction electron density is $n = 4 \; / \; a_{Cu}^3 = k_F^{3} \; / \; 3 \pi^2$, and $M(E_F) / A$ is given by the cross-section of the Fermi sphere (see Sec.~\ref{sec:mode-counting} above).  Applying Gauss' Theorem in Eq.~(\ref{eq:cond-ballist}), 
we derive 
\[
\frac{M(E_F)}{A} \; = \; \frac{\pi \; k_F^2}{4 \pi^2} = \frac{ (12^2 \cdot \pi)^{1/3}}{4 \; a_{Cu}^2} \; = \; 0.042 \; a_0^{-2}.
\]  

\subsection{Periodic Cu structures with disordered layers or vacancy scatterers \label{sec:disorder-vacancy}} 

\begin{figure}
(a) \includegraphics[scale=0.4]{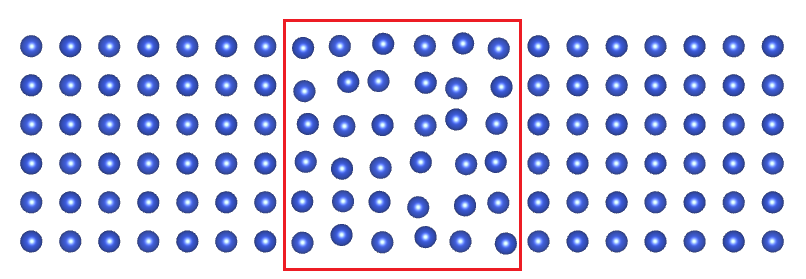} \\
(b) \includegraphics[scale=0.4]{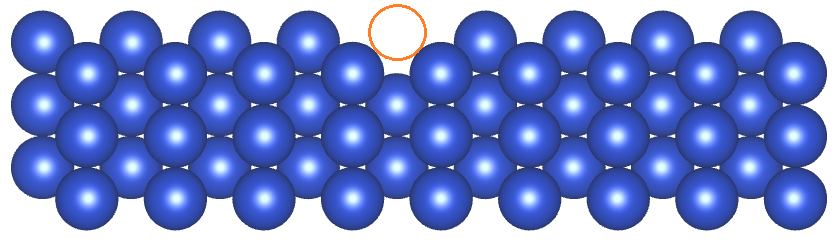}  \\
\caption{(a) Cu (100) bulk structure with 6 central monolayers having random disorder of RMS deviation 0.45 $a_0$, as in Ref.~\cite{Cu-GBs-Intel}.  Disordered layers are indicated by the box.  (b) Cu (100) bulk structure with central vacancies at density of one per 68.8 $a_0^2$ of cross-sectional area, as in Ref.~\cite{Cu-GBs-Intel}.  Vacancy is indicated by the empty circle.  Lateral boundary conditions are periodic.  \label{fig:disorder-vacancy}}
\end{figure}

We have also repeated transmission calculations in bulk (100) Cu structures with disordered layers or vacancies, using the same scattering region geometries as used previously in Ref.~\cite{Cu-GBs-Intel}.  In particular, the disordered structures had 3 or 6 central monolayers having random disorder of root mean square (RMS) deviation 0.45 $a_0$, and the vacancy structures had one vacancy per 68.8 $a_0^2$ of cross-sectional area.  
Fig.~\ref{fig:disorder-vacancy} shows the structures.  
For these calculations, we continued to use CAPs with $\Gamma_0 =$ 100 mRy and $\sigma =$ 8.5 $a_0$, as in Sec.~\ref{sec:bulk-Cu}.  Because of slight variations in the geometry or lattice constants, we repeated the ballistic calculations for the electrodes.  

We found reflection probabilities $R = T(E_F) / M(E_F) =$ 7\% for 3 disordered layers and $R = $ 13\% for 6 disordered layers, in fair agreement with Ref.~\cite{Cu-GBs-Intel}, which reported $R = $ 3\% and 12\%, respectively.  For the vacancy, we found $R = $ 14\%, in good agreement with Ref.~\cite{Cu-GBs-Intel}, which reported $R = $ 16\%.   

We consider these successful ballistic and scattering results in Cu to be a validation of the PBC method described in Sec.~\ref{sec:method-PBC}.  
In the remainder of this article, we determine and validate ABCs for use with 
bulk Rh electrodes.  
In our forthcoming work, we will use these validated ABCs to calculate reflection probability across Rh grain boundaries.

\section{Results for periodic Rh structures with bulk electrodes  \label{sec:Rh-results}}

\subsection{Bulk FCC Rh \label{sec:bulk-Rh}} 

Next, we performed 
simulations of ballistic conductance in bulk FCC Rh in the (100) and (111) orientations.  
For the Rh electrodes in (100) orientation, we used the same geometry as in the Cu bulk calculations of Sec.~\ref{sec:bulk-Cu}, 
but scaled the lattice constant to $a_{Rh} =$ 7.26 $a_0$ as optimized in a bulk primitive cell, and scaled all lengths and atomic coordinates accordingly.  Thus, the cell dimensions became 5.13 $a_0 \: \times$ 5.13 $a_0$ cross-section and 61.7 $a_0$ length from the first to the last monolayer, with a total cell length of 87 $a_0$ including vacuum.  We continued to use 0.37 $a_0$ grid spacing and 11 $\times$ 11 $k$-point sampling, resulting in $N =$ 33,700 grid points.  

For DFT, we used the local density approximation (LDA).  We used a norm-conserving Troullier-Martins pseudopotential for Rh with electronic configuration of 4$d^8$5$s^1$5$p^0$ and $s$/$p$/$d$ cutoff radii of 2.38 / 2.57 / 2.38 $a_0$.  For transmission, we used CAPs centered on the outermost monolayers, with strength $\Gamma_0$ in the range of 90 to 100 mRy and standard deviation $\sigma$ of 6.3 to 8.5 $a_0$, providing contact regions of at least 3.6 standard deviations for each CAP to decay before the central region.  
To explore the stability of our chosen CAP parameters, we also present $T(E)$ results using strength $\Gamma_0$ in the range of 50 to 120 mRy and standard deviation $\sigma$ of 4.8 to 12.0 $a_0$.  
We computed a fraction $p =$ 2\% of total complex eigenpairs, and tested this for convergence.  

\begin{figure}
(a) \includegraphics[scale=0.5]{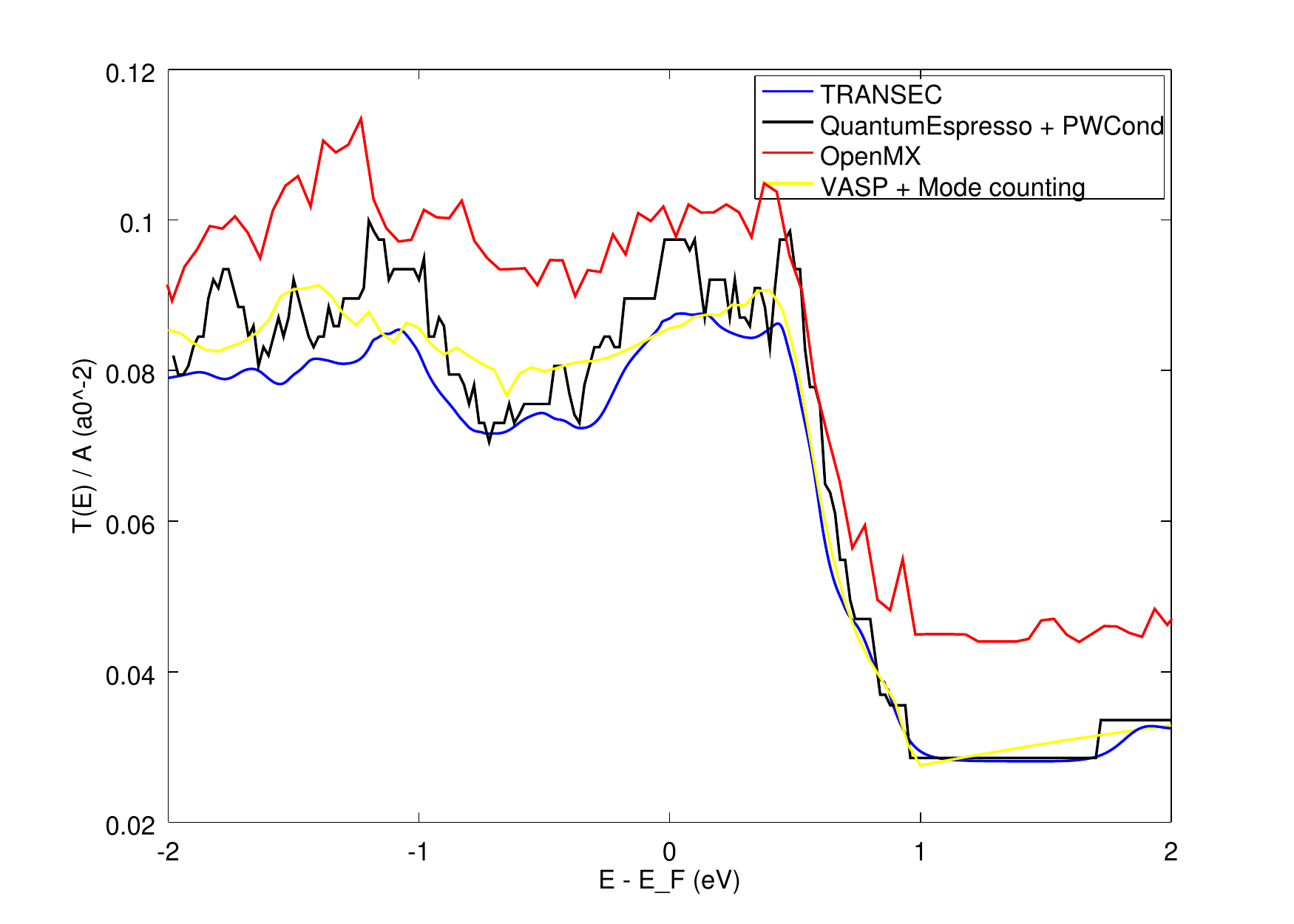}  \\
(b) \includegraphics[scale=0.5]{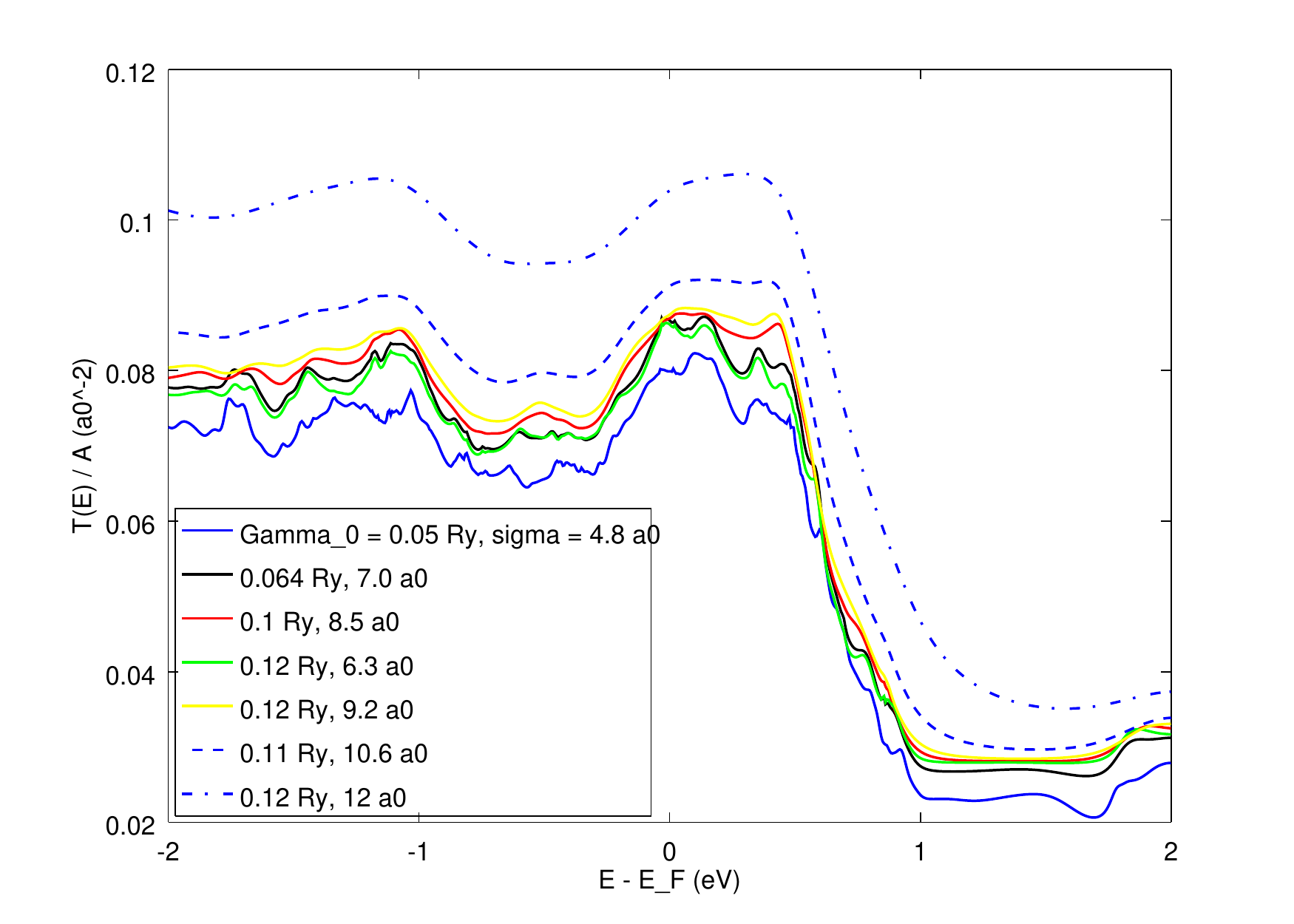} \\
\caption{(a) Rh (100) bulk transmission per unit area, $T(E) / A$, as a function of energy $E$.  Also shown for comparison are results from PWCOND, OpenMX, and our mode-counting method.  (b) Stability of $T(E)$ for CAP parameters $\Gamma_0 =$ 100 mRy, $\sigma =$ 8.5 $a_0$ compared with various other parameters. 
\label{fig:Rh-bulk-T-100}}
\end{figure}

Fig.~\ref{fig:Rh-bulk-T-100}(a) shows transmission for bulk FCC Rh in the (100) orientation.  Also shown for comparison are results from PWCOND, OpenMX, and our mode-counting method.  

Fig.~\ref{fig:Rh-bulk-T-100}(b) shows several $T(E)$ curves we generated with various CAP parameters.  
One way we validate our choice of CAPs is by determining a region of stability where $T(E)$ is relatively insensitive to the CAP parameters \cite{transec}.  As shown in Fig.~\ref{fig:Rh-bulk-T-100}(b), $T(E)$ is stable for CAPs in the region $\Gamma_0 =$ 0.1 Ry to 0.12 Ry, and $\sigma =$ 6,3 $a_0$ to 9.2 $a_0$.  
However, unlike the nanowire calculations reported in Ref.~\cite{transec}, 
here $T(E)$ displays insensitivity to $\Gamma_0$, yet scales monotonically over a range of $\sigma$ values.   
For example, the curves with $\Gamma_0 =$ 0.064 Ry, $\sigma =$ 7 $a_0$ and $\Gamma_0 =$ 0.12 Ry, $\sigma =$ 6.3 $a_0$ are more similar to each other than to the curve with $\Gamma_0 =$ 0.12 Ry, $\sigma =$ 12 $a_0$.

A second way we validate our choice of CAPs is by comparing $T(E)$ to accepted values and to calculations we performed using the mode counting method, as described in Sec.~\ref{sec:mode-counting} above.  
Fig.~\ref{fig:Rh-bulk-T-100}(a) shows results from OpenMX; PWCOND using electronic structure from Quantum Espresso; and our mode-counting method using electronic structure from VASP, which are in good agreement with TRANSEC.  

\begin{figure}
\includegraphics[scale=0.5]{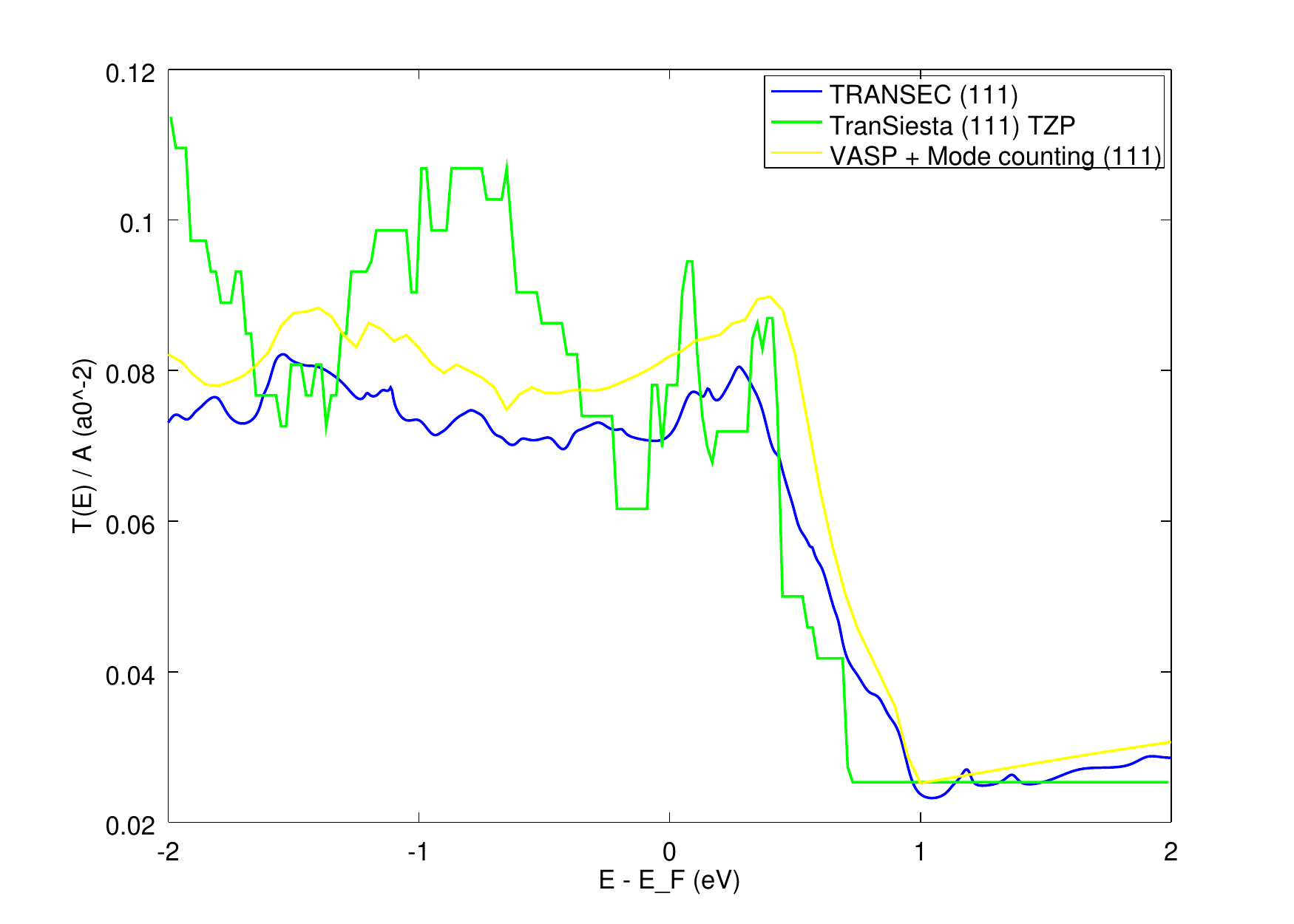}  
\caption{
Rh bulk transmission per unit area, $T(E) / A$, 
for the (111) orientation from TRANSEC, TranSiesta, and our mode-counting method.
\label{fig:Rh-bulk-T-111}}
\end{figure}

We also obtained $T(E)$ for the (111) orientation using TRANSEC, as well as TranSiesta, and our mode-counting method based on electronic structure from VASP.  The TRANSEC calculation used an orthogonal 9.7 $a_0 \; \times$ 8.4 $a_0$ unit cell, with a total of 134,000 grid points, and used the same CAP parameters as for the (100) orientation, $\Gamma_0 =$ 90 mRy and $\sigma =$ 6.3 $a_0$.  We performed the TranSiesta calculation with triple-zeta polarized (TZP) orbitals.  As shown in Fig.~\ref{fig:Rh-bulk-T-111}, these bulk calculations also agree.  

We find a ballistic conductance per unit cross-sectional area of $T(E_F)/A \;=
$ 0.087 $a_0^{-2}$, over twice that of Cu.  
While this result agrees with our independent calculations presented in Figs.~\ref{fig:Rh-bulk-T-100}(a) and \ref{fig:Rh-bulk-T-111}, 
note that it disagrees with results by Lanzillo, who found that Rh nanowires have only 14\% more ballistic conductance per unit cross-sectional area than Cu ones, even in a bulk-like regime where conductance scales linearly with wire cross-sectional area \cite{Lanzillo_2017}.  However, we 
performed multiple other validation checks of our results, as described here and in the next Section \footnote{
We do not present a Sommerfeld model of $G_B$ for Rh, as we did for Cu in Sec.~\ref{sec:bulk-Cu}, because Rh has a more complex Fermi surface and is not as readily modeled with a semi-classical effective mass approximation.  }.  
A possible explanation for this discrepancy is the precipitous drop from $T(E_F) =$ 0.087 $a_0^{-2}$ to $T(E_F \; + \; $1.1 eV$) =$ 0.028 $a_0^{-2}$.  Because linear-response conductance is governed by $T(E_F)$, a relatively small shift in the $T(E)$ curve along the energy axis could result in a significantly lower predicted conductance.  
For our PWCOND bulk Rh calculation, we found the $T(E)$ curve shown in Fig.~\ref{fig:Rh-bulk-T-100}(a) shifted to the left when the plane-wave cutoff energy was not converged, such that $T(E_F) / A \approx$ 0.06 $a_0^{-2}$.  
In addition, surface or other size effects in the Rh nanowire Lanzillo considers might contribute to this discrepancy, even though Lanzillo reports conductance scaling approximately linearly with cross-section.  

Finally, in view of the disagreement with results in Ref.~\cite{Lanzillo_2017} and the uncertainties mentioned above regarding the CAP region of stability, we wish to validate our CAPs further.  As a more strenuous validation of our CAPs, we have computed transmission in a large Rh point contact with periodic BCs and bulk electrodes, as described in the next Section.  

\subsection{Periodic Rh point contact with H as central atom \label{sec:point-contacts}} 

As further validation of the ABC parameters we identified for Rh (100) bulk electrodes, we computed $T(E)$ in a periodic Rh point contact having Rh (100) electrodes and H as a central ``device'' atom.  
This calculation also serves as a demonstration of the real-space PBC method in a large, periodic transition metal nanostructure without $R_{z,\pi}$ symmetry.  

As in Ref.~\cite{transec}, we rely on analytical understanding of the point contact transmission to validate 
results from TRANSEC.  
In particular, we choose H for the central atom since it has only a single electron, and so must have unit transmission at the Fermi level, provided it 
is isolated from the two electrodes and from its own periodic image.  
In choosing a central H atom, we thus circumvent 
any uncertainty in the value of $M(E_F)$ of the Rh electrodes.

\begin{figure}
\includegraphics[scale=0.4]{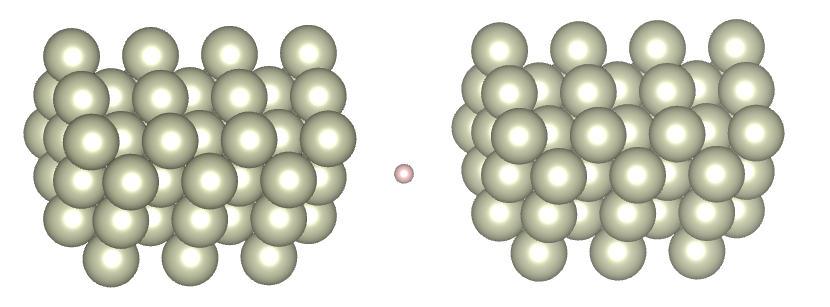} 
\caption{Structure of periodic Rh point contact with H as central atom, cross-section of 237.2 $a_0^2$, and electrode-central atom gap of 8.8 $a_0$.  Lateral boundary conditions are periodic. \label{fig:Rh-lg-point-contact-structure}}
\end{figure}

The geometry of this system is shown in Fig.~\ref{fig:Rh-lg-point-contact-structure}.  
For sufficiently large gaps between the electrodes and device atom, the central device atom's KS orbitals approach those of an isolated H atom.  
As a result, transmission near $E_F$ is 
dominated by the H atom's hybridized 1$s$ orbital, with a Lorentzian-shaped peak of height 1, and a width depending on the electrode-device gap. 
As shown, the central H atom is positioned slightly off-center, thereby breaking $R_{z,\pi}$ symmetry.  
Because we continued to use a ``slab'' geometry with periodic BCs in the lateral dimensions, 
far from the central region the electrodes comprised bulk (100) Rh, identical to those in Sec.~\ref{sec:bulk-Rh}.  Thus, we continued to use ABC parameters of height 100 mRy and standard deviation 8.5 $a_0$.  

Despite using bulk electrodes, in the limit of large lateral cross-section, we expect the transmission peak not to be modified by communication among the device atom and its periodic images -- a point we now address in greater depth.   
We find that when performing this calculation with a smaller 
cross-section 
of 10.26 $a_0 \: \times$ 10.26 $a_0$, 
the $k$-point-averaged peak $T(E \approx E_F)$ is about 20\% lower than expected, 
even though each \emph{individual} $k$-point resolved transmission curve $T_k(E)$ correctly displays a 
peak of unit height 
and approximately Lorentzian shape.  
Since the individual 
peaks do have height of 1, 
the low average 
is attributed to poor peak location alignment. 
We attribute this misalignment, 
in turn, to communication 
between the central H atom and its periodic images. 
In the limit of large cross-section, we anticipate that any electronic-structure band formed from different $k$-points should 
flatten and the individual $T_k(E)$ curves become identical, resulting in an averaged peak height of 1.  

\begin{figure}
(a)
\includegraphics[scale=0.5]{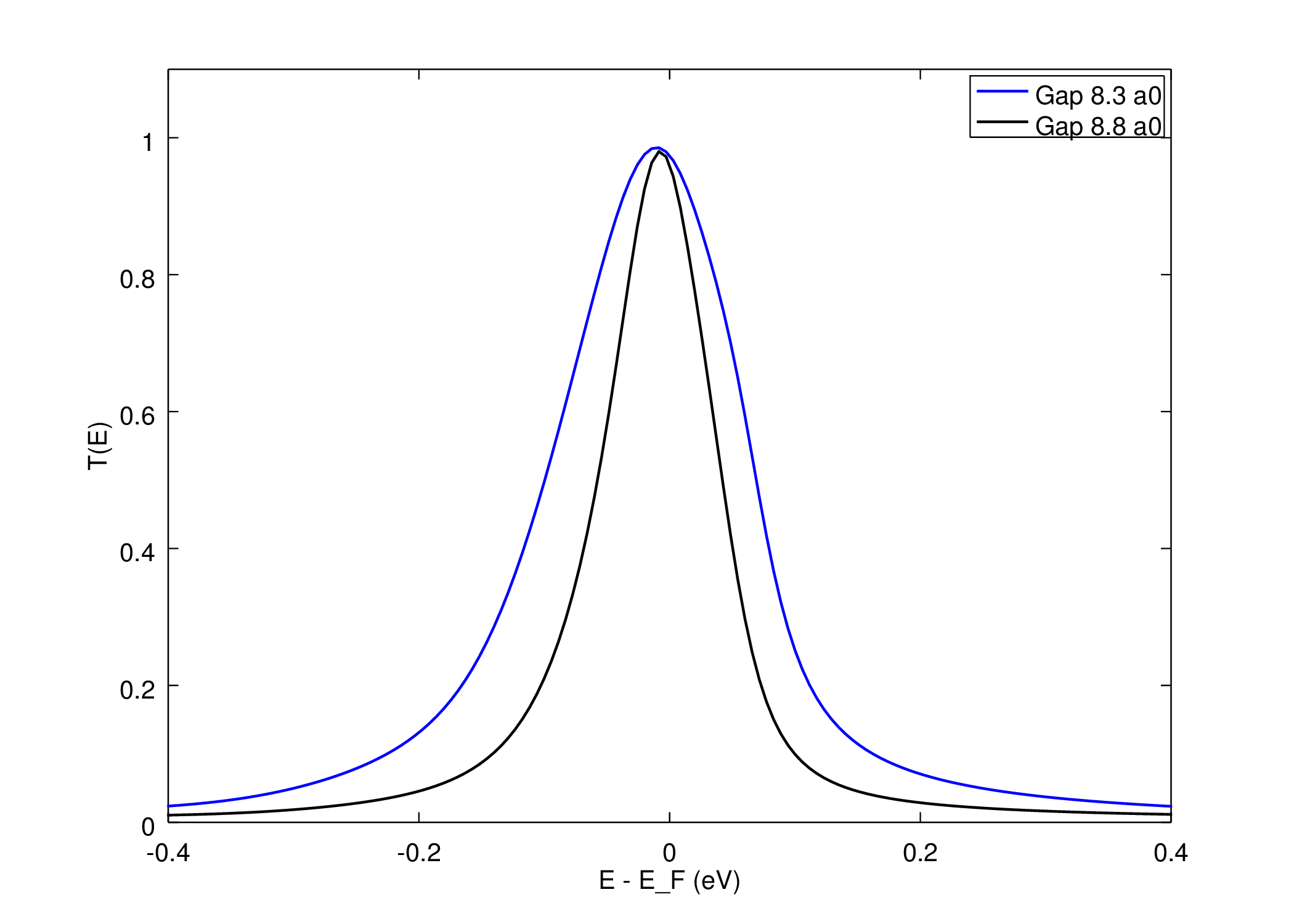} \\ 
(b)
\includegraphics[scale=0.5]{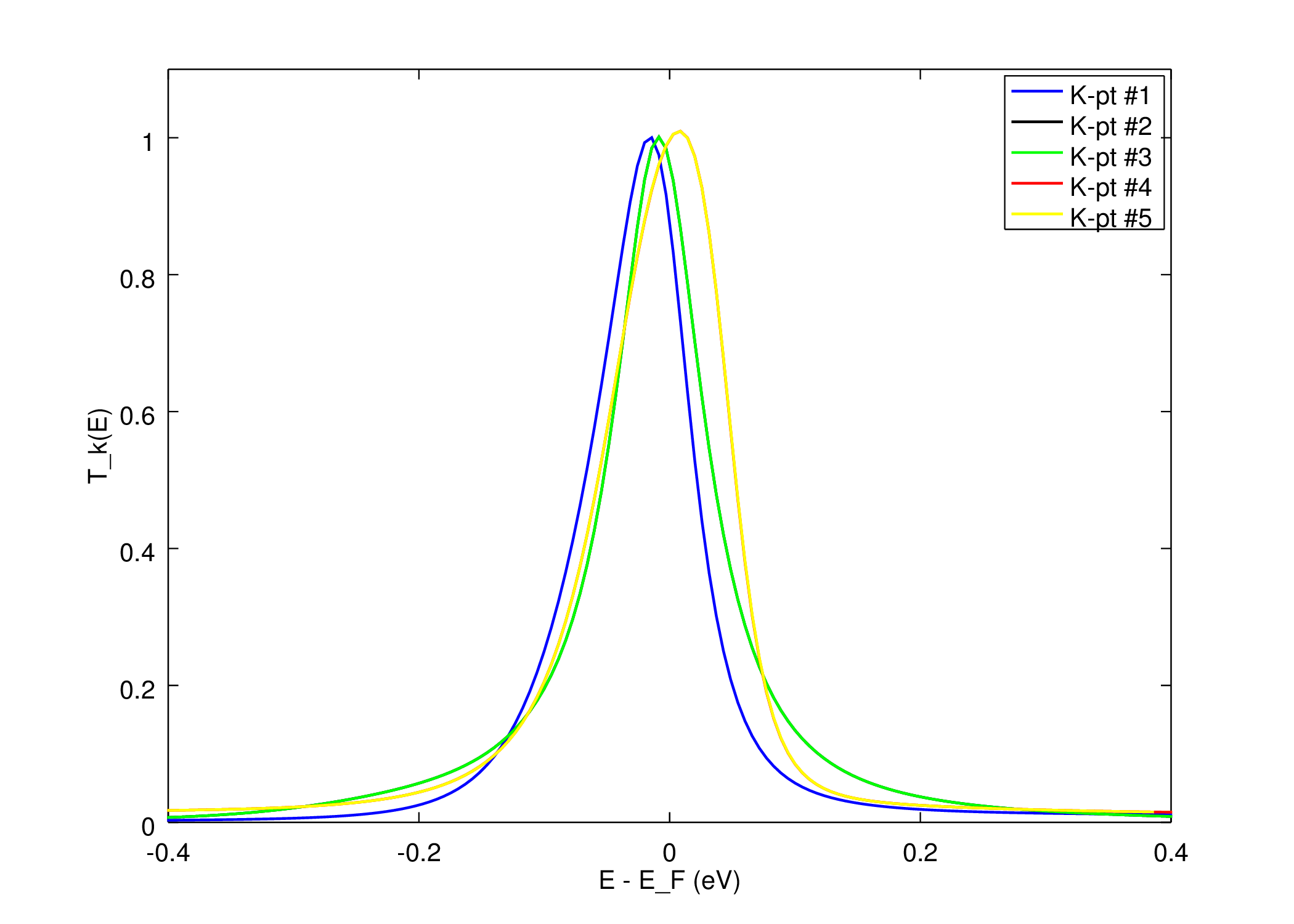} 
\caption{(a) Total transmission $T(E)$ for H atom-Rh electrode gaps of 8.3 $a_0$ and 8.8 $a_0$, and (b) $k$-point resolved transmission $T_k(E)$ as a function of energy $E$ for the large cross-section Rh point contact with H as central atom and H atom-Rh electrode gap of 8.8 $a_0$. 
The $k$-point resolved transmission $T_k(E)$ shows close peak-location alignment, such that the average $T(E)$ has a peak height of nearly 1 despite the individual $T_k(E)$ peaks being narrow.  
\label{fig:Rh-large-point-contact-T}}
\end{figure}

Therefore, we present a calculation of the point contact structure 
shown in Fig.~\ref{fig:Rh-lg-point-contact-structure}.   
For this calculation, the simulation cell had 3 $\times$ 3 of the tetragonal cells described in Sections \ref{sec:Cu-results} and \ref{sec:bulk-Rh}, 
for lateral dimensions of $ \frac{3 \sqrt{2}}{2} \; a_{Rh} \: \times \: \frac{3 \sqrt{2}}{2} \; a_{Rh} =$ 15.40 $a_0 \: \times$ 15.40 $a_0$ and a total cross-section of 237.2 $a_0^2$.  
Along the transport dimension, we used 7 monolayers, comprising 63 Rh atoms, for each electrode, providing more than 3.4 $\sigma$ to allow the CAPs to decay before the electrode-central atom gap \footnote{
We also tested for convergence of the electrode length.  We observed an odd-even effect, such that the average $T(E)$ for 6 monolayers differed slightly, but the result for 5 monolayers agreed well with the $T(E)$ result for 7 monolayers shown in Fig.~\ref{fig:Rh-large-point-contact-T}.  
}.   
The structure's length was thus approximately 60 $a_0$ from the first to the last monolayer, depending on the electrode-central atom gap, and the total cell length was 95.5 $a_0$ including vacuum.  
For this calculation, we used 0.385 $a_0$ grid spacing in order to reduce the computational burden, resulting in a total of $N =$ 358,100 grid points for the structure with shorter gap, and 361,000 grid points for the longer-gap structure.  
We used a 3 $\times$ 3 Monkhorst-Pack grid, for a total of 5 $k$-points after applying $R_{z, \pi}$ symmetry.  
For transport, we continued to use CAPs of height $\Gamma_0 =$ 90 mRy and standard deviation $\sigma =$ 6.3 $a_0$, in the region of stability, as discussed around Fig.~\ref{fig:Rh-bulk-T-100}(b) of the last section.

The results of this calculation 
are shown in Fig.~\ref{fig:Rh-large-point-contact-T}.  
Fig.~\ref{fig:Rh-large-point-contact-T}(a) shows the total transmission $T(E)$ for H atom-Rh electrode gaps of 8.3 $a_0$ and 8.8 $a_0$.  
As expected, the peak height is nearly 1, the peak location is approximately $E_F$, and the peak width narrows as the H atom-Rh electrode gap increases from 8.3 $a_0$ to 8.8 $a_0$.  
Fig.~\ref{fig:Rh-large-point-contact-T}(b) shows the $k$-point resolved transmission $T_k(E)$ for the calculation with electrode-central atom gap of 8.8 $a_0$.  As expected, the peak alignment of $T_k(E)$ is good, 
resulting in a peak height of nearly 1 in the average $T(E)$.

In some large point contact calculations, we also encountered difficulties converging $T_k(E)$ with respect to the number $p N$ of complex eigenpairs solved.  We attributed these difficulties to a loss of biorthogonality among separate eigenspaces, similar to the loss of orthogonality in Lanczos-like algorithms described by Paige's Theorem \cite{Parlett, eigensolution-ETH}.  Specifically, we observed the biorthogonal scalar product did not vanish between eigenvectors 
with relatively large eigenvalue separations, as it is expected to do.  
We found we could fix this issue by explicitly biorthogonalizing vectors in different eigenspaces.  
In response to this explicit biorthogonalization, the $T(E)$ curve converged much more reliably with $p N$.  
The results shown in Fig.~\ref{fig:Rh-large-point-contact-T} used $p N =$ 2,085 complex eigenpairs, and we tested this number for convergence.  

We also performed an even larger point contact calculation using a rectangular cross-section of 3 $\times$ 2 cubic cells, for a total cross-sectional area of 316.2 $a_0^2$, and a total of 500,000 grid points.  This structure contained 84 Rh atoms in each electrode, 
and used a 2 $\times$ 3 Monkhorst-Pack grid.  This calculation also displayed good alignment of the $k$-point resolved peaks $T_k(E)$, an averaged $T(E)$ peak of approximately 1, and a peak width depending on the electrode-central atom gap, as expected analytically.   

\section{Conclusions}

We have described and validated a method to include periodic boundary conditions in real-space calculations of electronic conductance.  We demonstrated this method in bulk Cu and Rh as well as large Rh point contacts with H as the central atom.  We developed CAP parameters for bulk Rh electrodes, to be used for Rh grain boundary simulations in subsequent work.  Although the resulting bulk Rh $T(E)$ disagrees with results in Ref.~\cite{Lanzillo_2017}, we have validated our CAP parameters against numerous calculations with other conductance packages, as well as analytical expectations for the point contact structure.  

\section{Acknowledgements}

We are grateful for the use of computing resources at the TIAL group in the UW Electrical \& Computer Engineering Department.  
BF would like to thank Oded Hod, Amir Natan, and Leeor Kronik for helpful discussions.

\appendix*
\section{Notes on implementation and parallelization \label{sec:implementation}}

We have implemented the PBC methods of Sec.~\ref{sec:method-PBC} in TRANSEC by performing each $k$-point calculation independently.  In addition, we have implemented collinear spin polarization, which performs the calculation for each spin $\sigma$ independently (but in this work we present only unpolarized calculations of non-magnetic materials).  
Because the $k$-points and spins are independent, the individual $T_{k, \sigma}(E)$ calculations for spin $\sigma$ and lateral Bloch vector $k$ can be performed on separate computing nodes with {\em distributed} memory.  Hence, the $T_{k, \sigma}(E)$ calculations can be naturally parallelized over $k$ and $\sigma$, or can be performed in serial on the same nodes.  

However, a drawback to parallelization is that each $T_{k, \sigma}(E)$ calculation requires its own independent memory, 
driving up the total memory requirement.  The number of parallelization groups should therefore be chosen to balance the available memory and computing time resources.  
Alternatively, the total memory and/or computing time requirements can be reduced by partitioning the eigenspectrum of individual $T_{k, \sigma}(E)$ calculations, as described in Ref.~\cite{Transec-partitions}.  

For the case with $R_{z,\pi}$ symmetry, the eigenvectors for $+\vec{k}$ and $-\vec{k}$ points are closely related, so our calculations are restricted to those $k$-points with non-negative $k_a$, 
reducing the computational burden by about half.  For the case without $R_{z,\pi}$ symmetry, separate eigensolutions must be performed for both $+\vec{k}$ and $-\vec{k}$ points.  But as Eqs.~(\ref{eq:transform-CAP-4})-(\ref{eq:transform-no-symm}) show, the eigenpairs for both $+\vec{k}$ and $-\vec{k}$ points must be combined to determine $T_{k}(E)$.  As a result, we prefer to perform $+\vec{k}$ and $-\vec{k}$ eigensolutions in serial in the same parallelization group, so as to avoid additional communication and load-balancing issues when combining the 
eigenpairs.  
However, we have implemented automated parallelization of the $+\vec{k}$ and $-\vec{k}$ eigensolutions in the relatively rare cases when the calculation as a whole can be performed in less elapsed time this way.  

The computed eigenpairs for $+\vec{k}$ are then matched to those for $-\vec{k}$ by determining the best agreement between respective eigenvalues in these two sets,  
while discarding or otherwise resolving spurious or missing eigenpairs that have no sufficiently close match.  
We find the matching eigenvalues generally agree to about $10^{-5}$ of the next-best match.  
For the $\Gamma$-point, i.e. $\vec{k} =$ 0, even without $R_{z,\pi}$ symmetry it is only necessary to perform eigensolution once, since $\vec{k} = -\vec{k}$, so if practical we balance the computing load by assigning a greater share of $k-$points to the group with the $\Gamma$-point.

\bibliographystyle{unsrt} 
\bibliography{PBCs_Rh_Bulk_PointContact}

\end{document}